\documentclass[aps,pra,showpacs]{revtex4}
\usepackage{amsmath}
\usepackage{graphicx}
\usepackage{dcolumn}
\usepackage{longtable}
\begin{document}

% version submitted to the arXiv

\title{Cosmological perturbations on local systems}

\author{Gregory S. Adkins}
\email[]{gadkins@fandm.edu}
\author{Jordan McDonnell}
\affiliation{Department of Physics, Franklin and Marshall College, Lancaster, Pennsylvania 17604}
\author{Richard N. Fell}
\affiliation{Department of Physics, Brandeis University, Waltham, Massachusetts 01742}

\date{\today}

\begin{abstract}
We study the effect of cosmological expansion on orbits--galactic, planetary, or atomic--subject to an inverse-square force law.  We obtain the laws of motion for gravitational or electrical interactions from general relativity--in particular, we find the gravitational field of a mass distribution in an expanding universe  by applying perturbation theory to the Robertson-Walker metric.  Cosmological expansion induces an ($\ddot a/a) \vec r$ force where $a(t)$ is the cosmological scale factor.  In a locally Newtonian framework, we show that the $(\ddot a/a) \vec r$ term represents the effect of a continuous distribution of cosmological material in Hubble flow, and that the total force on an object, due to the cosmological material plus the matter perturbation, can be represented as the negative gradient of a gravitational potential whose source is the material actually present.  We also consider the effect on local dynamics of the cosmological constant.  We calculate the perihelion precession of elliptical orbits due to the cosmological constant induced force, and work out a generalized virial relation applicable to gravitationally bound clusters.
\end{abstract}

\pacs{04.90.+e, 98.80.Jk}

\maketitle

%%%%%%%%%%%%%%%%%%%%%%%%%%%%%%%%%%%%%%%%%%%%%%%%%%%

\section{Introduction}

If the universe has expanded by a factor of $1000$ since the first atoms were formed, and by $30$ some percent since the formation of the solar system, it seems reasonable that there must be {\it some} effect, even if small, of this expansion on the orbits of atoms and planets.  On the other hand, spacetime is locally flat, and first-order deviations from flatness are accounted for by Newtonian gravitational theory, leaving no room for stray cosmological effects on local systems.  So, do atomic or planetary orbits respond at least a bit to the universal expansion or don't they?  Our purpose in this article is to answer this question both within matter only cosmological models and in the currently accepted benchmark model including a cosmological-constant derived dark energy component.

Many authors have studied the effects of cosmological expansion on local dynamics with various and often conflicting conclusions.  Most of the early work weighs against any cosmological effect on planetary orbits.  A seminal contribution was made by McVittie \cite{McVittie33}, who found an exact solution to Einstein's equation that he interpreted as representing a point mass  embedded in a Robertson-Walker expanding universe.  McVittie, and later J\"arnefelt \cite{Jarnefelt40,Jarnefelt42} found that planetary orbits in such a model do not participate in the cosmic expansion.  Recent work on problems and progress in the interpretation of McVittie's solution is summarized by Nolan \cite{Nolan99}, who considers the influence of the cosmic expansion on planetary orbits in the McVittie universe still to be an open question.  In another influential study, Einstein and Straus \cite{Einstein45} (see also Sch\"ucking \cite{Schucking54}) showed that a truncated Schwarzschild solution could be embedded in a background expanding universe with smoothly matched boundary conditions.  In this picture, orbits around the central mass are completely unaffected by the expansion.  However, this approach has been criticized on various counts, summarized by Carrera and Giulini \cite{Carrera06}.  More recently, Dicke and Peebles argued on general grounds that there are no local effects \cite{Dicke64}.  Their work was criticized by Noerdlinger and Petrosian \cite{Noerdlinger71} and by Carrera and Giulini \cite{Carrera06} on the grounds that they neglected the ``friction'' term in their equation that would have produced a positive effect.  Anderson \cite{Anderson95} obtained an equation of motion including such a friction term, but found that there are no expansion-induced effects on circular orbits, at least in the leading order of approximation.  On the other side, many authors find small but non-vanishing effects on local orbits due to cosmological expansion. \cite{Carrera06, Noerdlinger71, Pachner64, Gautreau84, Cooperstock98, Bonnor99, Dominguez01, Bolen01, Price05}  We will discuss many of these results in the course of this work.

We base our approach on relativistic cosmology, and show that locally this reduces to a Newtonian framework.  For local systems this is an appropriate approximation, as has been discussed by McCrea and Milne \cite{McCrea34, McCrea55} and particularly by Dicke and Peebles \cite{Dicke64} and Callan, Dicke, and Peebles \cite{Callan65} for the present application.  We restrict our attention to a spatially flat universe, which is in accord with recent observations.  In any case, a flat space approximation will be adequate for space scales much less than the radius of curvature.  We consider models involving matter and dark energy only, and neglect radiation as of negligible import in the present epoch.

%%%%%%%%%%%%%%%%%%%%%%%%%%%%%%%%%%%%%%%%%%%%%%%%%%%

\section{Equations of motion including cosmological effects}

In this section we derive the equation for non-relativistic motion of a point particle in the electric or gravitational field of a spherically symmetric static source.  The equation has the same form for either case:
\begin{equation} \label{nr_eqn_of_motion}
\ddot {\vec r} = -\frac{\alpha}{r^2} \hat r + \frac{\ddot a}{a} \vec r
\end{equation}
where $\vec r $ is the position of the point particle in physical coordinates, $\alpha$ represents the strength of the interaction, and $a(t)$ is the cosmological scale factor.

The provenance of (\ref{nr_eqn_of_motion}) is quite different for the two cases.  For electric attraction, it comes from the curved space generalization of the Lorentz force law
\begin{equation} \label{Lorentz_force_law}
\ddot x^\mu + \Gamma^\mu_{\alpha \beta} \dot x^\alpha \dot x^\beta = \frac{q}{m} F^\mu_{\; \; \nu} \, \dot x^\nu \, ,
\end{equation}
where $\Gamma^\mu_{\alpha \beta}$ are the Christoffel coefficients, $F^{\mu \nu}$ is the electromagnetic field tensor, $x^\mu$ is the particle coordinate in co-moving coordinates (so $\vec r = a(t) \vec x$), and the dot denotes a derivative with respect to proper time.  On the other hand, for gravitational attraction, the equation of motion is the geodesic equation
\begin{equation}
\ddot x^\mu + \Gamma^\mu_{\alpha \beta} \dot x^\alpha \dot x^\beta = 0
\end{equation}
with no separate force term--the effect of the source is contained in $\Gamma^\mu_{\alpha \beta}$.  Since the electric case is more straightforward, we will consider it first.

The Robertson-Walker metric for a flat expanding spacetime has the form
\begin{equation}
g_{\mu \nu} = \begin{pmatrix} 1 & 0 \\ 0 & -a^2(t) \delta_{i j} \end{pmatrix} \, ,
\end{equation}
where $a(t)$ satisfies the cosmological Einstein equations
\begin{eqnarray}
\frac{3 \dot a^2}{a^2} &=& 8 \pi G \bar \rho \, , \nonumber \\
-\frac{\dot a^2}{a^2} - 2 \frac{\ddot a}{a} &=& 8 \pi G \bar p \, , \label{cosmological_Einstein_eqns}
\end{eqnarray}
in terms of the total cosmological proper density $\bar \rho$ and pressure $\bar p$.
The corresponding Christoffel coefficients
\begin{equation}
\Gamma^\mu_{\alpha \beta} = \frac{1}{2} g^{\mu \nu} \left ( g_{\nu \alpha, \beta} + g_{\nu \beta, \alpha} - g_{\alpha \beta , \nu} \right )
\end{equation}
are $\Gamma^0_{i j} = a \dot a \delta_{i j}$, $\Gamma^i_{0 j} = \Gamma^i_{j 0} = (\dot a/a) \delta_{i j}$, with all others vanishing.  The field tensor is
\begin{equation}
F_{\mu \nu} = \begin{pmatrix} 0 & \tilde E_i \\ -\tilde E_i & -\epsilon_{i j k} \tilde B_k \end{pmatrix}
\end{equation}
where $\tilde E_i$ and $\tilde B_i$ are related to the electric and magnetic fields.  The field tensor satisfies Maxwell's equations
\begin{eqnarray}
F^{\mu \nu}_{\; \; \; \; ; \nu} = 4 \pi j^\mu \, , \\
F_{\mu \nu , \alpha} + F_{\nu \alpha , \mu} + F_{\alpha \mu , \nu} = 0 \, .
\end{eqnarray}
For a spherically symmetric static source, Maxwell's equations require $\tilde B_i=0$ and
\begin{equation}
\tilde E^i = \frac{C}{a(t) x^2} \hat r_i
\end{equation}
outside the source, where $x=\sqrt{x_1^2+x_2^2+x_3^2}$, $\hat r = \vec x/x$, and $C$ is some constant.  Now the spatial components of the Lorentz force law (\ref{Lorentz_force_law}) lead to
\begin{equation}
\ddot x^i + 2 \frac{\dot a}{a} \dot x^i \dot x^0 = \frac{q}{m} \frac{C}{a^3 x^2} \hat r_i \dot x^0 \, .
\end{equation}
For non-relativistic motion this is
\begin{equation}
\ddot x^i + 2 \frac{\dot a}{a} \dot x^i = \frac{q}{m} \frac{C}{a^3 x^2} \hat r_i \, ,
\end{equation}
and in terms of the physical coordinates $\vec r = a(t) \vec x$ it takes the form
\begin{equation} \label{elec_eq_of_motion}
\ddot r_i - \frac{\ddot a}{a} r_i = \frac{Q q}{m} \frac{\hat r_i}{r^2}
\end{equation}
since we can now identify $C$ with the source charge $Q$.

We derive the equation of motion for a test mass in a gravitational field in two steps.  First, we use the Einstein equation to find the metric for an expanding universe containing a localized non-relativistic distribution of matter in addition to the uniformly distributed cosmological material.  We restrict ourselves to small distortions from the cosmological background.  After obtaining the metric we use the geodesic equation in the weak-field limit to obtain the equation of motion.

A massive object such as the sun inserted into a background expanding universe perturbs the usual Robertson-Walker metric.  Exact solutions describing inhomogeneous cosmologies have been extensively studied (reviewed for example in the book of Krasi\'nski \cite{Krasinski97}), but for our purposes a perturbative approach will be adequate physically and more straightforward to interpret.
Accordingly, we will consider perturbations to the Robertson-Walker metric. \cite{Irvine65, Kodama84, Mukhanov92, Dodelson03}  For the weak-field case with only non-relativistic motion of localized source material it is sufficient to consider the scalar perturbation
\begin{equation}
g_{\mu \nu} = \begin{pmatrix} 1+2 \psi(t,\vec x \,) & 0 \\ 0 & -(1-2 \chi(t, \vec x \,) ) a^2(t) \delta_{i j}
\end{pmatrix}
\end{equation}
and a stress-energy tensor of the form
\begin{equation}
T_{\mu \nu} = \sum_a \left [ (\bar \rho_a+\bar p_a) U_\mu U_\nu - \bar p_a g_{\mu \nu} \right ] + \rho u_\mu u_\nu
\end{equation}
where $\bar \rho_a$ and $\bar p_a$ are the proper density and pressure of the $a^{\rm th}$ component of cosmological material (matter:$\, m$, dark energy:$\, \Lambda$) and $\rho(t,\vec x \,)$ is the proper density of the matter perturbation.  To emphasize, $\bar \rho_m+\rho$ represents all of the matter actually present, so $\rho$ itself is the actual (total) matter minus the cosmological matter.  The four-velocity $U_\mu$ has only a time component and is normalized: $U^\mu U_\mu=1$; while $u_\mu \approx U_\mu$ for non-relativistic motion of the perturbing matter.  The perturbative nature of this ansatz is signaled by the small size of $\psi$ and $\chi$.  The global behavior of the metric is governed by the cosmological parameters $\bar \rho_a$ and $\bar p_a$, so that even in regions where the perturbation $\rho$ is larger than the cosmological density $\bar \rho$ the effect of the perturbation is just to change the metric by a small amount: $\psi, \, \chi << 1$. 

Einstein's equation has the form
\begin{equation}
G_{\mu \nu} = -8 \pi G T_{\mu \nu}
\end{equation}
where $G_{\mu \nu}=R_{\mu \nu}-(1/2) g_{\mu \nu} R$ is the Einstein tensor.  The spatial off-diagonal components of $T_{\mu \nu}$ are small--of $O(v^2)$ where $v<<1$ is a typical non-relativistic speed of the matter in the perturbation $\rho$.  With the neglect of such small contributions, the Einstein equation implies $\partial_i \partial_j \psi = \partial_i \partial_j \chi$ for all $i$ and $j$.  This and the vanishing of $\psi$ and $\chi$ at large distances leads to the requirement $\psi=\chi$.  The form of the time-space off-diagonal elements of $G$ ($G_{0 i} = 2 \partial_i(\partial_0 \psi + \frac{\dot a}{a} \psi)$) suggests that the substitution $\psi \rightarrow \phi/a$ would be helpful, so in fact the metric that we use has the form
\begin{equation} \label{grav_metric}
g_{\mu \nu} = 
\begin{pmatrix} 1+ \frac{2 \phi(t,\vec r \, )}{a(t)} & 0 \\ 0 & - \left ( 1 - \frac{2 \phi(t,\vec r \, )}{a(t)} \right ) 
a^2(t) \delta_{i j} 
\end{pmatrix} \, .
\end{equation} 
The Einstein tensor for this metric, through first order in the weak field $\phi$, is \cite{endnote1}
\begin{equation}
G_{\mu \nu} = 
\begin{pmatrix} -\frac{3 \dot a^2}{a^2} \left ( 1+\frac{2 \phi}{a} \right )
- \frac{2}{a^3} \nabla^2 \phi + \frac{6 \dot a}{a^2} \partial_0 \phi & -\frac{2}{a} \partial_0 \partial_i \phi \\ 
- \frac{2}{a} \partial_0 \partial_i \phi &  \left [\dot a^2 + 2 a \ddot a \left ( 1-\frac{3 \phi}{a} \right ) 
- 4 \dot a \partial_0 \phi - 2 a \partial_0^2 \phi \right ] \delta_{i j} 
\end{pmatrix} \, .
\end{equation} 
We are constrained to work to zeroth order in the flow velocity $v$ of the perturbing mass distribution since our metric is not sufficiently complex to respond to the flow except through the time dependence of
$\phi(t,\vec x\,)$.  The metric perturbation $\phi$ can be obtained from the time-time component of the Einstein equation, which is
\begin{equation}
\frac{3 \dot a^2}{a^2} \left ( 1+\frac{2 \phi}{a} \right )+ \frac{2}{a^3} \nabla_x^2 \phi -
\frac{6 \dot a}{a^2} \partial_0 \phi =
8 \pi G \left ( \bar \rho U_0^2 + \rho u_0^2  \right )  \, .
\end{equation}
Here $\bar \rho = \bar \rho_m+\bar \rho_\Lambda$ represent the total cosmological energy density.  We neglect the term involving the time derivative of $\phi$ since $\partial_0 \phi$ is of the order of the flow velocity.  Also we neglect $\dot a^2 \phi$ compared to $\nabla_x^2 \phi$.  Now using 
$u_0^2 \approx U_0^2 = g_{00} = 1+\frac{2 \phi}{a}$ and the cosmological equations
(\ref{cosmological_Einstein_eqns}), the equation for $\phi$ reduces to
\begin{equation}
\nabla_x^2 \phi = 4 \pi G a^3 \rho(t, \vec x \,) \, .
\end{equation}
The ``potential'' $\phi$ has the usual solution
\begin{equation}
\phi(t,\vec x\,) = -G a^3 \int \frac{d^3 x' \, \rho(t,\vec x\,')}{\vert \vec x - \vec x\,' \vert}
= -G \int \frac{dM'}{\vert \vec x - \vec x\,' \vert}
\end{equation}
since $\rho$ is the proper mass density
\begin{equation}
\rho(t,\vec x\,) = \frac{dM}{d^3 r} \, .
\end{equation}

Our next task is to obtain the equation of motion for a point test mass moving in the gravitational field created by the mass distribution $\rho$ in an expanding universe.  The equation of motion is the geodesic equation
\begin{equation}
\ddot x^\mu + \Gamma^\mu_{\alpha \beta} \dot x^\alpha \dot a^\beta = 0 \, ,
\end{equation}
where the Chrostoffel coefficients are calculated from the metric of (\ref{grav_metric}).  They are
\begin{eqnarray}
\Gamma^0_{0 0} &=& \frac{1}{a} \left ( -\frac{\dot a}{a} \phi + \partial_0 \phi \right ) \, , \cr
\Gamma^0_{0 i} = \Gamma^0_{i 0} &=& \frac{\partial_i \phi}{a} \, , \cr
\Gamma^0_{i j} &=& a \left ( \dot a - 3 \frac{\dot a}{a} \phi - \partial_0 \phi \right )  \delta_{i j} \, , \cr
\Gamma^i_{0 0} &=& \frac{\partial_i \phi}{a^3} \, , \cr
\Gamma^i_{0 j} = \Gamma^i_{j 0} &=& \frac{1}{a} \left (\dot a + \frac{\dot a}{a} \phi - \partial_0 \phi \right )  \delta_{i j} \, , \cr
\Gamma^i_{j k} &=& -\frac{1}{a} \left ( \delta_{i k} \partial_j \phi + \delta_{i j} \partial_k \phi
- \delta_{j k} \partial_i \phi \right ) \, .
\end{eqnarray}
Then the equation of motion, in the non-relativistic and weak-source approximations $\dot x^0 \approx 1$, $\vert \dot {\vec x} \vert <<1$, $\phi/a <<1$, and $\partial_0 \phi<<1$, has the form
\begin{equation}
\ddot x^i + 2 \frac{\dot a}{a} \dot x^i = -\frac{\partial_i \phi}{a^3} \, .
\end{equation}
Proceeding as for the electric case, the equation of motion in physical coordinates is
\begin{equation} \label{grav_eq_of_motion}
\ddot {\vec r} - \frac{\ddot a}{a} \vec r = - \vec \nabla_r \Phi \, .
\end{equation}
The physical-coordinate potential $\Phi(t, \vec r \,) = \phi(t, \vec x \,)/a$ satisfies
\begin{equation} \label{phys_pot_eqn}
\nabla_r^2 \Phi = 4 \pi G \tilde \rho(t, \vec r \,)
\end{equation}
where $\tilde \rho(t, \vec r \,) = \frac{dM}{d^3 r} = \frac{1}{a^3} \frac{dM}{d^3 x} = \frac{1}{a^3} \rho(t, \vec x \,)$ is the proper density of the matter perturbation in physical coordinates.  For a stationary spherically symmetric mass distribution the potential is $\Phi = -G M/r$ and (\ref{grav_eq_of_motion}) has the same form as the electric field equation (\ref{elec_eq_of_motion}).

We note here that (\ref{phys_pot_eqn}), giving an instantaneous connection between mass density here and potential there, will break down unless the size of our local region ($R$) is small enough so that the cosmological mass density doesn't change much in the time $R/c$ necessary for information to propagate across the region.  In other words, we require $R << R_H$ where $R_H=c/H_0$ is the Hubble distance.

%%%%%%%%%%%%%%%%%%%%%%%%%%%%%%%%%%%%%%%%%%%%%%%%%%%

\section{Radial expansion of orbits as a mathematical problem} 

In this section we will consider the effect of the $\ddot a/a$ cosmological force term on orbits as a purely mathematical problem--discussion of the physical content of this equation will be delayed until the next section.  Two kinds of distortions of Keplerian orbits  have been discussed in the literature: changes to the orbital radius and cosmologically induced precession.  In this section we will focus on the cosmological effect on the radius.

In order to make sense of the literature, it is essential to realize that two types of radius change have been considered.  These are, first, the change with time of the radius of a circular orbit found for example by studying the position of the minimum of the effective potential; and second, the change with time of the radius of an initially circular orbit found by solving the equations of motion.  We pause to make the distinction more clearly.

The first type of radius change can be described more completely as follows.  Let the value of the cosmological force term at some initial time $t_1$ be $f_1$.  With the net force consisting of the force due to the central mass (or charge) plus $f_1$, and with a given angular momentum, there is a unique radius $r_1$ for a circular orbit.  The particle will never actually trace out this orbit as long as the cosmological force term is changing, but this is the orbit that it {\it would} have were the cosmological force constant.  At a later time $t_2$ the cosmological force has value $f_2$.  At this time, with the {\it same} angular momentum as at time $t_1$, there is a different radius $r_2$ for a circular orbit with the new force.  This definition of the radius change $\delta r = r_2-r_1$ was used by Gautreau \cite{Gautreau84}, Cooperstock, Faraoni and Vollick \cite{Cooperstock98}, and by Bonnor \cite{Bonnor99} in his Appendix B.  The most common usage was to calculate the radius change per orbit by multiplying the time rate of change of $\delta r$ by the orbital period.

The second type of radius change was found by starting a particle on what would be a circular orbit (were the cosmological force constant) of radius $r_0$ at time $t_0$ and following the trajectory of the particle through one revolution.  In this approach, $\delta r$ is the difference between the final radius and $r_0$.  More specifically, the particle is launched with the initial conditions $r(t_0)=r_0$, $\dot r(t_0)=0$, and $\ddot r(t_0)=0$, where $r_0$ and the angular momentum are such that, were the cosmological force constant at its value at time $t_0$, the orbit would be circular.  The particle follows the trajectory required by the equation of motion, and ends up after one revolution at radius $r_0+\delta r$.  This is what Bonnor \cite{Bonnor99} calculates  in the body of his paper.  Price \cite{Price05} also looked at orbits found as solutions to the equation of motion.  This solution to the equation of motion tells where the particle actually is, as opposed to the approximate position resulting from the first approach.

We point out that a particle obeying the equations of motion in a time varying potential will not necessarily remain at the minimum of the potential energy curve if it starts there, but for a slowly varying potential the particle will remain at least near the minimum.  This behavior will be seen for particle orbits in an expanding universe.

In the remainder of this section we will give explicit calculations of the two types of radius change for a flat, dust-filled universe.  We use the non-relativistic equation of motion
\begin{equation} \label{vec_eq_of_motion}
\ddot {\vec r} = - \frac{\alpha}{r^2} \hat r + \frac{\ddot a}{a} \vec r
\end{equation}
where $\alpha = G M$ for motion under the influence of gravitation around a central mass $M$; and $\alpha = -Q q/m$ for motion of a charge $q$ with mass $m$ around a central charge $Q$.  For a flat dust-filled universe, the scale factor is $a(t) = (t/t_0)^{2/3}$ where $t_0$ is the age of the universe, so $\ddot a/a = -2/(9 t^2)$.  In spherical coordinates, (\ref{vec_eq_of_motion}) implies angular momentum conservation
\begin{equation}
h = r^2 \dot \phi = const.
\end{equation}
(where over-dots now indicate time derivatives), and the radial equation
\begin{equation} \label{eq_of_motion}
\ddot r - \frac{h^2}{r^3} + \frac{\alpha}{r^2} + \frac{2}{9 t^2} r = 0 \, .
\end{equation}

\subsection{Radius of a circular orbit}

Circular orbits have $\ddot r=0$, so the radius is given by the solution to
\begin{equation} \label{circ_orbit}
-\frac{h^2}{r^3} + \frac{\alpha}{r^2} + \frac{2}{9 t^2} r = 0 \, .
\end{equation}
For very large time the cosmological force in this model vanishes, and for a given value of $h$ we call the corresponding circular-orbit radius $r_0$.  We find $r_0 = h^2/\alpha$.  At finite times the circular-orbit radius $r$ satisfies (\ref{circ_orbit}) and can be written as $r=r_0 (1+\delta)$.  The value of $r$ is very close to $r_0$ since the age of the universe $t$ is so much larger than the orbital period $\tau = 2 \pi r^2/h$, which is the only other time scale in the problem.
We rewrite (\ref{circ_orbit}) as
\begin{equation}
h^2 \delta = -\frac{2 r_0^4}{9 t^2} (1+\delta)^4 \, ,
\end{equation}
and find that approximately $\delta \approx -2 r_0^4/(9 h^2 t^2)$.  So for the orbital radius we have
\begin{equation} \label{orbital_radius}
r \approx r_0 \left ( 1 - \frac{1}{18 \pi^2} \left ( \frac{\tau}{t} \right )^2 \right ) \, .
\end{equation}
This is precisely result (4.10) of Cooperstock \textit{et al.}  \cite{Cooperstock98}.  It is also the result one would obtain by following the position of the minimum of the effective potential $U(r) = h^2/(2 r^2) - h^2/(r_0 r) + r^2/(9 t^2)$ for fixed $h$.  The effective potential can be written as
\begin{equation}
U(r) = \frac{h^2}{r_0^2} \left \{ \frac{1}{2 \xi^2} - \frac{1}{\xi} + \frac{\xi^2}{36 \pi^2} \left ( \frac{\tau}{t} \right )^2 \right \}
\end{equation}
where $\xi = r/r_0$.  The minimum is at $\xi=1$ when $t \rightarrow \infty$, but is slightly smaller than one for finite $t$.  The position of the minimum grows according to (\ref{orbital_radius}) as $t$ increases.
The increase of radius (\ref{orbital_radius}) during the course of one period, evaluated at the present cosmological time $t_0$, is
\begin{equation} \label{circular_result}
\delta r \approx \frac{dr}{dt} \tau = \frac{r_0}{9 \pi^2} \left ( \frac{\tau}{t_0} \right )^3 \, .
\end{equation}
This is equivalent to the results (6.26) of Gautreau \cite{Gautreau84} and (B4) of Bonnor \cite{Bonnor99}, but differs from (59) of Carrera and Guilini \cite{Carrera06,endnote2}.   

Of course, the result (\ref{orbital_radius}) for $r(t)$ is not where the particle will actually be if free to move in the specified field of force.  Instead, we have found the radius at time $t$ of a circular orbit with a given angular momentum and with the cosmological force held artificially fixed.

\subsection{Orbital out-spiraling}

The second type of cosmologically induced radial perturbation that has been considered is that found by solving the equation of motion (\ref{vec_eq_of_motion}) for a particle moving on an initially-circular orbit.  We choose initial conditions to ensure that at the present moment (time $t_0$), were the cosmological force to remain fixed, the particle would move in a circular orbit of radius $r_0$.  That is, the initial conditions are $r(t_0)=r_0$, $\dot r(t_0)=0$, and $\ddot r(t_0)=0$.  The time parameter $t$ has an awkward zero point for our purposes here, so we reparameterize time according to
\begin{equation}
t=t_0 (1+s) \, ,
\end{equation}
where the new time parameter $s$ is small for times $t$ not so different from the initial time $t_0$.  We will use a power series approach to solve the equation of motion, so we write the radial variable as
\begin{equation}
r = r_0 \left ( 1 + \sum_{n=1}^\infty a_n s^n \right ) \equiv r_0 \kappa(s) \, .
\end{equation}
The initial conditions immediately tell us that $a_1=a_2=0$.  In fact, the vanishing of $\ddot r(t_0)$ and the radial equation of motion (\ref{eq_of_motion}) give a relation among $r_0$, $h$, and $\alpha$:
\begin{equation} \label{ddotr_condition}
-\frac{h^2}{r_0^3} + \frac{\alpha}{r_0^2} + \frac{2}{9 t_0^2} r_0 = 0 \, .
\end{equation}
In terms of the new parameters, and with use of (\ref{ddotr_condition}), the radial equation of motion (\ref{eq_of_motion}) takes the form
\begin{equation} \label{eq_of_motion_2}
\frac{d^2 \kappa}{ds^2} + \Lambda^2 \frac{\kappa-1}{\kappa^3} + \frac{2}{9} \left ( \frac{\kappa}{(1+s)^2} - \frac{1}{\kappa^2} \right ) = 0
\end{equation}
where $\Lambda=h t_0/r_0^2 \approx 2 \pi t_0/\tau$ is a very large parameter.  We find the values of the $a_n$ by equating the coefficients of the various powers $s^n$, and obtain the series solution for $\kappa(s)$:
\begin{equation} \label{kappa_series}
\kappa(s) = 1 + \frac{2}{27} s^3 - \frac{1}{18} s^4 + \left ( \frac{17}{405} - \frac{\Lambda^2}{270} \right ) s^5 + \left ( -\frac{253}{7290} + \frac{\Lambda^2}{540} \right ) s^6 + O(s^7) \, .
\end{equation}
This series may be truncated at the $s^3$ term as long as $s^3 << \Lambda^2 s^5$; that is, for times $\tilde t = t-t_0=t_0 s$ small compared to an orbital period.  For times $\tilde t$ comparable to an orbital period, $\Lambda s$ is of order one, and the leading contributions for all coefficients $a_n$ with $n$ odd are of order $\Lambda^{n-3} s^n = O(\Lambda^{-3})$.  The next-to-leading terms for the odd coefficients have order $\Lambda^{-5}$, while the even coefficients have main contributions of order 
$\Lambda^{-4}$.  This allows us to sum the principal contributions for $\tilde t=O(\tau)$ and get the approximate analytic form
\begin{eqnarray}
\kappa(s) = 1 + \frac{4}{9} \left ( \frac{s^3}{3!} - \frac{\Lambda^2 s^5}{5!} + \frac{\Lambda^4 s^7}{7!} + \cdots \right ) + O \left ( \frac{1}{\Lambda^4} \right ) \cr
= 1+\frac{4}{9 \Lambda^3} \left ( \Lambda s - \sin (\Lambda s) \right ) + O \left ( \frac{1}{\Lambda^4} \right )\, .
\end{eqnarray}
The secular term increases linearly with time with the same coefficient
\begin{equation}
\frac{dr}{d t} \approx \frac{r_0}{9 \pi^2} \frac{\tau^2}{t_0^3}
\end{equation}
as in (\ref{circular_result}).  So the actual radius follows the minimum of the effective potential as time progresses but with some sloshing around represented by the $\sin(2 \pi \tilde t/\tau)$ term.  This is in accord with expectations based on the adiabatic theorem of classical mechanics \cite{Landau81}.

Bonnor \cite{Bonnor99} calculated the out-spiral radius increase starting from the relativistic equation of motion using a power series approach.  His results (47) and thus (51) agree with our power series solution for $\kappa(s)$ through the term of order $s^3$.  However, in order to apply the series expansion for values of time as large as an orbital period, an infinite subset of terms must be summed, as shown above.  We see that special relativistic kinematics do not affect the result to this order (as was suggested in \cite{Bonnor99}), since our calculation was purely non-relativistic.  

%%%%%%%%%%%%%%%%%%%%%%%%%%%%%%%%%%%%%%%%%%%%%%%%%%%

\section{Physical effects of cosmic expansion}

Now we come to the real question: is there a cosmologically induced perturbation to planetary and Coulombic orbits, or not?  In order to approach this question, we begin by considering the physical interpretation of the $\frac{\ddot a}{a} \vec r$ force.  If the physical picture specifies the superposition of a mass $M$ on a background homogeneous expanding universe, then besides the usual gravitational attraction to $M$ there will be an additional gravitational attraction from the cosmological mass $\frac{4}{3} \pi r^3 \bar \rho_m$ inside the planet's orbit.  The radial force due to this mass is $f_r = -G (\frac{4}{3} \pi r^3 \bar \rho_m) \frac{1}{r^2} = - \frac{4 \pi G}{3} \bar \rho_m r$.  This mass is continually decreasing on account of the outflow of cosmological mass through a spherical surface of radius $r$ due to the Hubble flow, which leads to a decreasing $f_r$.  The connection of $f_r$ to the scale factor is through the Friedmann equations
\begin{eqnarray}
\frac{3 \dot a^2}{a^2} = 8 \pi G \bar \rho \, , \\
-\frac{\dot a^2}{a^2} - 2 \frac{\ddot a}{a} = 8 \pi G \bar p \, ,
\end{eqnarray}
where $\bar \rho$ and $\bar p$ are the total cosmological mass density and pressure.  If we focus on a matter-only universe for the moment, $\bar p=\bar p_m=0$ and $\frac{\ddot a}{a} = -\frac{4 \pi G}{3} \bar \rho_m$.  The cosmological force $f_r = \frac{\ddot a}{a} r$ in the equation of motion is present exactly to represent the gravitation of the cosmological mass.  This agrees with the interpretation of Gautreau \cite{Gautreau84}.

Our position then is as follows.  If the model includes the cosmological mass spread homogeneously throughout the universe and following Hubble's expansion law, then cosmological perturbations to orbits will in fact occur.  Otherwise there will be no perturbations to the orbits.  It is imperative that on the scale set by the size of the system, it must be a good approximation to consider the cosmological matter to be continuous.  An example of a model of this nature is Bonnor's hydrogen atom situated in the Einstein-de Sitter (flat, dust-only) universe with the proton co-moving with the cosmological fluid \cite{Bonnor99}.  The dust must be composed of parts small compared to the size of the atom, and must be homogeneous as seen on a scale set by the size of the atom.  In such a world, the electron (considered classically, of course) would spiral outwards according to the calculations of the preceding section.

On the other hand, if the model considered does not include actual cosmological mass expanding according to the Hubble law and homogeneous on a scale set by the size of the system, then there are no cosmological perturbations.  This is the situation, for example, in the solar system.  It is not possible to imagine a mass like the sun simply placed in a background cosmology without affecting the cosmological mass located nearby.  For example, consider the sun and earth as a binary system.  With the origin at the center of the sun, the Hubble-law speed for cosmological mass at the location of the earth is only $3.4 \times 10^{-7} m/s$, and it would take just $57 \mu s$ for the sun to bring this mass to rest and initiate its plunge into the sun.  Clearly a model of the solar system including cosmological Hubble flow is not physically reasonable.

Now recall that the perturbation is based on the difference between the actual mass present and the cosmological mass.  So the $-\vec \nabla \Phi$ force in (\ref{grav_eq_of_motion}) is the gravitational force due to the perturbative mass.  The cosmological force,  $\frac{\ddot a}{a} \vec r$, also has an associated potential $-\frac{\ddot a}{a} \frac{r^2}{2}$ so the potential due to the total mass (cosmological plus perturbation) is 
$\tilde \Phi = - \frac{\ddot a}{a} \frac{r^2}{2} + \Phi$.  It follows that the force law can be rewritten as
%\begin{equation}
%\ddot {\vec r} = - \vec \nabla \tilde \Phi \, ,
%\end{equation}
\begin{equation}
\ddot {\vec r} = \frac{Q q}{m} \frac{\hat r}{r^2} - \vec \nabla \tilde \Phi \, ,
\end{equation}
where the electric term accounts for the electric force, if any.  The explicit cosmological $(\ddot a/a) \vec r$ force term has disappeared: all gravitational forces are due to the mass actually present.  If there is in fact cosmological mass present in Hubble flow, its effect is included in the $- \vec \nabla \tilde \Phi$ force.

The effect of dark energy is apparently a different story.  There is no known clumpiness to dark energy: it seems to be a property of the vacuum itself, and is thus truly homogeneous.  The universe at present seems to be well-described by a two-component model consisting of cosmological-constant dark energy and matter.  Using $\bar \rho_\Lambda$ and $\bar p_\Lambda=-\bar \rho_\Lambda$ for the dark energy density and pressure, and $\bar \rho_m$ for the matter density (and $\bar p_m=0$), the Friedmann equations become
\begin{eqnarray}
\frac{3 \dot a^2}{a^2} = 8 \pi G (\bar \rho_m + \bar \rho_\Lambda) \, , \\
-\frac{\dot a^2}{a^2} - 2 \frac{\ddot a}{a} = 8 \pi G \bar p_\Lambda \, .
\end{eqnarray}
It follows that the cosmological force is proportional to
\begin{equation}
\frac{\ddot a}{a} = -\frac{4}{3} \pi G \bar \rho_m + \frac{8}{3} \pi G \bar \rho_\Lambda \, .
\end{equation}
The $\bar \rho_m$ term in $\ddot a/a$ combines with the perturbation to give the force due to the actual, physical mass distribution.  The dark energy term can be written as $H_0^2 \Omega_\Lambda$ where $H_0$ is the present value of the Hubble constant and $\Omega_\Lambda$ measures the density of dark energy in terms of the critical density: $\bar \rho_\Lambda = \Omega_\Lambda \rho_c$ where $\rho_c = \frac{3 H_0^2}{8 \pi G}$.  For values we use $H_0 = 70(7) (km/s)/Mpc = 2.27(0.23) \times 10^{-18} s^{-1}$ and $\Omega_\Lambda = 0.7$. \cite{Ryden03}  Including the dark energy effect (but no charges), the force equation takes the form
\begin{equation} \label{eq_of_motion_final}
\ddot {\vec r} = - \vec \nabla \tilde \Phi  + k \vec r \, ,
\end{equation}
where again $\tilde \Phi$ is the potential due to the physically present matter, and $k \vec r$ is the linear repulsive force due to the cosmological constant with $k = H_0^2 \Omega_\Lambda$.

The linear cosmological repulsion is a real force and is potentially observable.  The challenge of course is the very small size of the coefficient $k$: $k \approx 3.6 \times 10^{-36} s^{-2}$.  The $k \vec r$ force is time independent, so there will be no spiraling of planetary orbits.  The orbits will just be slightly larger than they would be without the presence of this force.  As a modification of the $1/r^2$ gravitational force law, the $k \vec r$ force will cause precession of elliptical orbits.  We have calculated the rate of this precession for orbits of period $\tau$ and eccentricity $e$, and find
\begin{equation} \label{precession_1}
\Delta \theta_p = \frac{3 k}{4 \pi} \tau^2 \sqrt{1-e^2}
\end{equation}
radians per orbit. \cite{endnote2a} This agrees with the result of Kerr, Hauck, and Mashhoon. 
\cite{Kerr03}  For orbits in the solar system this is too small to be observed. \cite{endnote3}
An alternate expression for (\ref{precession_1}) is
\begin{equation} \label{precession_2}
\Delta \theta_p \approx 0.16 \left ( \frac{\tau}{t_0} \right )^2 \sqrt{1-e^2} \, ,
\end{equation}
which tells us that the precession is only large for systems with periods an appreciable fraction of the age of the universe.  Excessively short human lifetime precludes any effective application of (\ref{precession_2}) to real orbits.

A potentially useful application of the cosmological repulsion is as a correction to the virial relation for galaxy clusters.  The ``moment of inertia'' of a cluster can be defined to be
\begin{equation}
I = \frac{1}{2} \sum_a M_a \vec r_a \cdot \vec r_a \, ,
\end{equation}
where $a$ labels the constituents of the cluster.  For a cluster in steady-state, the moment of inertia is constant.  This fact, combined with the equation of motion (\ref{eq_of_motion_final}), yields
\begin{eqnarray}
0 = \ddot I = \sum_a M_a \left ( \dot {\vec r}_a \cdot \dot {\vec r}_a + \vec r_a \cdot \ddot {\vec r}_a \right ) \\
= M \langle v^2 \rangle - \frac{\alpha G M^2}{R_{1/2}} + 2 I k \, ,
\end{eqnarray}
where $M$ is the mass of the cluster, $R_{1/2}$ is the radius containing half of the cluster mass, and $\alpha$ is a constant of order one defined by $\sum_{a,b} \frac{M_a M_b}{\vert {\vec r}_b - {\vec r}_a \vert} = \frac{2 \alpha M^2}{R_{1/2}}$.  We define an additional constant $\beta$ of order one by $2 I = M \langle r^2 \rangle = \beta M R_{1/2}^2$, and find that
\begin{equation} \label{virial_formula}
\frac{\alpha G M}{R_{1/2}} = \langle v^2 \rangle + \beta k R_{1/2}^2
\end{equation}
as the generalization of the usual virial formula. \cite{endnote4}
As an example of the use of (\ref{virial_formula}), we evaluate the effect of the cosmological repulsion for the case of the Coma cluster discussed in Ryden Sec.~8.3 \cite{Ryden03}.  The mean square speed has the (somewhat model dependent) value $\langle v^2 \rangle = 2.32 \times 10^{12} (m/s)^2$.  If we assume an exponentially falling mass distribution, the cosmological repulsion term contributes $\beta k R_{1/2}^2 \approx (1.68) (3.6 \times 10^{-36}) (4.6 \times 10^{22} m)^2 \approx 1.3 \times 10^{10} (m/s)^2$.  This corrects the mass estimate by about a half of one percent, and is negligible considering the sizes of the other uncertainties.  However, the cosmological repulsion becomes more important for more extended clusters, and would make a contribution comparable to that of the gravitational term for clusters that are only marginally bound.

Related work includes that of Lavav et al. \cite{Lahav91}, who used a cosmological constant enhanced virial relation to discuss galaxy formation.  Current simulations of structure formation using the $\Lambda$CDM model include the effect of the cosmological constant.

%%%%%%%%%%%%%%%%%%%%%%%%%%%%%%%%%%%%%%%%%%%%%%%%%%%

\section{Conclusion}

Motion in local systems is governed by the gravitational forces due to the actual mass present, along with other forces such as electromagnetism, the linearly repulsive cosmological constant force, and initial conditions.  There is no additional effect caused by cosmological expansion: the $(\ddot a/a) \vec r$ cosmological force combines with the gravitational force due to the perturbation of the mass distribution (mass present above and beyond the cosmological mass) to produce the gravitational force due to the distribution of mass actually there.  It follows that calculations like those of Sec.~3, while interesting, do not correspond to physically realizable situations for gravitational systems.  It is possible in principle to imagine two charged objects with one in orbit around the other, along with a continuum (at least relative to the size of the orbit) of out-flowing cosmological dust, for which the orbit spiraling of Sec.~3 would in fact occur, but there are no known systems of this type in practice.  On the other hand, if ``dark energy'' is homogeneous with a constant density, the $k \vec r$ dark energy force is real and will affect particle orbits.  We have calculated the perihelion precession due to the $k \vec r$ force, and have also found the generalized form of the virial formula relating mean-square speed, mass, and size for bound clusters of gravitating objects.  These effects are marginally important for large systems such as clusters of galaxies.

%%%%%%%%%%%%%%%%%%%%%%%%%%%%%%%%%%%%%%%%%%%%%%%%%%%

\begin{acknowledgments}
We are grateful to Howard Schnitzer for useful comments on the manuscript.  We acknowledge the support of the Hackman Scholars Program of Franklin \& Marshall College.
\end{acknowledgments}

%%%%%%%%%%%%%%%%%%%%%%%%%%%%%%%%%%%%%%%%%%%%%%%%%%%


\begin{thebibliography}{10}
\bibitem{endnote1} The Christoffel symbols and Einstein tensor were calculated with the aid of the Mathematica notebook supplied on the web site for the text Gravity by J. B. Hartle \cite{Hartle03}.  We adjusted conventions in the notebook to agree with the timelike metric that we are using.
\bibitem{endnote2}  Carrera and Giulini \cite{Carrera06} seem to assume (incorrectly) that the fractional change $\Delta x/x$ in the co-moving radius of an orbit is the same as the fractional change $\delta r/r$ in the physical radius (in the paragraph below their (59)), from which they conclude that $\delta r/r$ is of the order of $\tau/t_H$.  Such a result for $\delta r/r$ would be unacceptably large.
\bibitem{endnote2a}  This is a first order perturbation theory result, and as such simply adds to the usual general relativistic precession of $43$ arcsec per century.  Were the cosmological constant larger this precession could be comparable with or even larger than the GR result.  The fact that these two precessions simply add (to first order) is clear from the form of the full relativistic equation of orbital motion around a mass $M$ in a space with a non-zero cosmological constant (see Rindler \cite{Rindler01}).  In this equation of motion the post-Newtonian effects of relativity and of $\Lambda$ are additive.
\bibitem{endnote3} For mercury, with eccentricity $e=0.206$, semimajor axis $a_{\rm sm}=5.79 \times 10^{10}m$, and period $\tau = 0.241 y = 7.60 \times 10^6 s$ \cite{Tholen00}, the predicted precession is  $\Delta \theta_p = 4.9 \times 10^{-23} \,{\rm rad/orbit} = 4.2 \times 10^{-15} \, {\rm arcsec/ century}$.  This is much smaller than the observational uncertainty of about 
$0.005 \, {\rm arcsec/ century}$ \cite{Pitjeva05,Iorio06}, and places no useful bound on the size of the cosmological constant (despite the claim by Cardona and Tejeiro \cite{Cardona98}, and in agreement with the observation by Wright \cite{Wright98} and others).
\bibitem{endnote4} Some typical values for $\alpha$ and $\beta$ are as follows.  For a spherical distribution of constant mass density and finite radius: $\alpha=0.48$ and $\beta=0.95$.  For an exponential distribution $\rho(r)=A \exp(-\lambda r)$: $\alpha=0.42$ and $\beta=1.68$.
\end{thebibliography}
\end{document}